\documentclass[nofootinbib,pra,aps,twocolumn,showpacs,showkeys,amsmath,amssymb,superscriptaddress,final,reprint,floatfix,longbibliography]{revtex4-1}
\usepackage[latin9]{inputenc}
\usepackage{color}
\usepackage{amstext}
\usepackage{amssymb}
\usepackage{graphicx}
\definecolor{mygrey}{gray}{0.35}
\definecolor{myblue}{rgb}{0.0,0.1,0.8}
\definecolor{myzard}{cmyk}{0,0,0.05,0}
\definecolor{mywhite}{rgb}{1,1,1}
\definecolor{myred}{rgb}{1,0.,0.3}

\usepackage[colorlinks=true,citecolor=myblue,linkcolor=myred,urlcolor=myblue,breaklinks=false]{hyperref}

\makeatletter
\@ifundefined{textcolor}{}
{%
 \definecolor{BLACK}{gray}{0}
 \definecolor{WHITE}{gray}{1}
 \definecolor{RED}{rgb}{1,0,0}
 \definecolor{GREEN}{rgb}{0,1,0}
 \definecolor{BLUE}{rgb}{0,0,1}
 \definecolor{CYAN}{cmyk}{1,0,0,0}
 \definecolor{MAGENTA}{cmyk}{0,1,0,0}
 \definecolor{YELLOW}{cmyk}{0,0,1,0}
}

\usepackage{amsmath}
\usepackage{graphicx}
\usepackage{amssymb}
\usepackage{txfonts}
\usepackage{subfigure}
\usepackage{epstopdf}

\newcommand{\ket}[1]{\left\vert #1 \right\rangle}
\newcommand{\bra}[1]{\left\langle #1 \right\vert}
\newcommand{\ketbra}[2]{\ket{ #1}\bra{ #2} }
\newcommand{\bla}[1]{\left( #1 \right)}
\newcommand{\blb}[1]{\left[ #1 \right]}

\makeatother

\begin{document}

\title{Unambiguous nuclear spin detection using engineered quantum sensing sequence}
\author{Zijun Shu}
\author{Zhendong Zhang}
\author{Qingyun Cao}
\author{Pengcheng Yang}
\affiliation{School of Physics, Huazhong University of Science and Technology, Wuhan 430074, China}
\affiliation{International Joint Laboratory on Quantum Sensing and Quantum Metrology, Huazhong University of Science and Technology, Wuhan, China, 430074}
\author{Martin B. Plenio}
\affiliation{Institut f\"{u}r Theoretische Physik $\&$ IQST, Albert-Einstein Allee 11, Universit\"{a}t Ulm, D-89081 Ulm, Germany}
\affiliation{International Joint Laboratory on Quantum Sensing and Quantum Metrology, Huazhong University of Science and Technology, Wuhan, China, 430074}
\author{Christoph M\"{u}ller}
\author{Johannes Lang}
\author{Nikolas Tomek}
\author{Boris Naydenov}
\author{Liam P. McGuinness}
\affiliation{Institut f\"{u}r Quantenoptik $\&$ IQST, Albert-Einstein Allee 11, Universit\"{a}t Ulm, D-89081 Ulm, Germany}
\author{Fedor Jelezko}
\affiliation{Institut f\"{u}r Quantenoptik $\&$ IQST, Albert-Einstein Allee 11, Universit\"{a}t Ulm, D-89081 Ulm, Germany}
\affiliation{International Joint Laboratory on Quantum Sensing and Quantum Metrology, Huazhong University of Science and Technology, Wuhan, China, 430074}
\author{Jianming Cai}
\email{jianmingcai@hust.edu.cn}
\affiliation{School of Physics, Huazhong University of Science and Technology, Wuhan 430074, China}
\affiliation{International Joint Laboratory on Quantum Sensing and Quantum Metrology, Huazhong University of Science and Technology, Wuhan, China, 430074}

\begin{abstract}
Sensing, localising and identifying individual nuclear spins or frequency components of a signal in the presence of a noisy environments requires
the development of robust and selective methods of dynamical decoupling.
An important challenge that remains to be addressed in this context are
spurious higher order resonances in current dynamical decoupling sequences
as they can lead to the misidentification of nuclei or of different
frequency components of external signals. Here we
overcome this challenge with engineered quantum sensing sequences that
achieve both, enhanced robustness and the simultaneous suppression of
higher order harmonic resonances. We demonstrate experimentally the
principle using a single nitrogen-vacancy center spin sensor which
we apply to the unambiguous detection of external protons.
\end{abstract}


\date{\today}

\maketitle

{\it Introduction.---} The detection of single nuclear spins represents an important yet challenging step towards single molecule magnetic resonance spectroscopy and imaging \cite{Rugar04} which holds the promise for the observation of individual protein structures without the need for crystallization of large ensembles. The achievement of this goal may have significant impact on structural biology and medical imaging and may also provide a new tool for the investigation of nuclear spin dynamics in non-trivial quantum biological processes \cite{Hue13,Lambert13}. Recently, research has made impressive progress in single nuclear spin detection using various physical platforms \cite{Vincent12,Pla13,Zhao12,Kol12,Tam12}, particularly in the highly challenging task of detecting single spins in external molecules \cite{Muller14,Mamin13,Stau13,Sus14,Shi15,Per14}. Among those physical platforms, a
sensor based on individual negatively charged nitrogen-vacancy (NV) centers
in diamond \cite{Doherty13,Wu16} has demonstrated appealing prospects for applications in biology and medicine, due to its biocompatibility, nano-scale size and long coherence
times under ambient conditions \cite{Schir14,Erm13}. NV centers, shallowly implanted
to within a few nanometers below the diamond surface, enable strong coupling between
NV sensors and target nuclei, thereby promoting the detection sensitivity
from a relative large ensemble of nuclei \cite{Mamin13,Stau13} to single nuclear spin
sensitivity in small clusters of nuclear spins \cite{Muller14,Sus14,Shi15,Per14}. Besides
the interest in single molecule magnetic resonance spectroscopy and imaging, single nuclear
spin addressing also has an important role to play in the precise coherent
control of nuclear spin qubits, where it may facilitate the realisation of quantum
memories with long coherence times and nuclear spin based quantum information processors \cite{Jiang09,Maurer12,Waldherr14,Cai13}.

One ultimate goal of single molecule magnetic resonance spectroscopy using a single
spin sensor is to detect a single nuclear spin and further infer the structure of a
single molecule \cite{CaiNJP13}. This would generally require a larger number of
dynamical sensing pulses\cite{Lukasz08} such that even tiny pulse errors may accumulate
and deteriorate significantly the magnetic resonance spectroscopy signal \cite{Zhao14,Ma15,Greiner15,Albrecht15,Cas15,Wang15}. Furthermore, the effect of a
finite pulse duration, during which the control pulses and the interaction with the
nuclei/field simultaneously act on the sensor, leads to spurious higher order resonances
in the measurement data that may lead to a misidentification of frequency components
in the signal. Indeed, as a prominent recent example these spurious resonances have
recently been identified as the source for the misidentification
of nuclei in work towards nuclear magnetic resonance spectroscopy of biological systems
\cite{Loretz15}. In particular, the ratio of gyromagnetic moments
$\mu_{^1\mbox{H}}/\mu_{^{13}\mbox{C}} \simeq 4$ of hydrogen ($^1 \mbox{H}$) and carbon
($^{13} \mbox{C}$) which leads to the overlap of different resonances preventing
their unique attribution to specific nuclei \cite{Loretz15}. The presence of $^{13}
\mbox{C}$ as part of the target molecule or the diamond itself is difficult to avoid
making this an urgent problem to address \cite{Sta15}.

In this work, we both theoretically and experimentally demonstrate quantum sensing pulse sequences with engineered phases that effectively suppress spurious higher order resonances and are, at the same time, robust to pulse errors. In particular, we find that a type of  MLEV-8 pulse sequence \cite{Lev83} denoted as YY8 sequence $\bla{\frac{\pi}{2}}_{\phi_i}- \blb{\bla{\pi}_{-y} \bla{\pi}_{y} \bla{\pi}_{y} \bla{\pi}_{-y}\bla{\pi}_{-y} \bla{\pi}_{-y}\bla{\pi}_{y} \bla{\pi}_{y}}^{N}-\bla{\frac{3\pi}{2}}_{\phi_i}$ as a representative example can eliminate unwanted higher order spurious harmonic response in nuclear magnetic resonance spectroscopy and additionally can exhibit enhanced robustness. Using single nitrogen-vacancy spin sensor in diamond, we demonstrate the simultaneous suppression of both the 2$^{nd}$ and the 4$^{th}$ order spurious resonance e.g. from the surrounding $^{13} \mbox{C}$ nuclei in diamond in proton sensing. We also observe the effect of the initial phase $\phi_i$ in the XY8 family pulse sequences $\bla{\frac{\pi}{2}}_{\phi_i}- \blb{\bla{\pi}_{x} \bla{\pi}_{y} \bla{\pi}_{x} \bla{\pi}_{y}\bla{\pi}_{y} \bla{\pi}_{x}\bla{\pi}_{y} \bla{\pi}_{x}}^{N}-\bla{\frac{3\pi}{2}}_{\phi_i}$ \cite{Haase16}. By choosing appropriate initial phase $\phi_i$, either the 2$^{nd}$ or the 4$^{th}$ order spurious resonance may be suppressed individually (though not simultaneously in contrast with YY8 pulse sequence). Based on engineered quantum sensing pulse sequences that suppress spurious signals, we detect the hydrogen signal in an unambiguous way using shallow implanted NV center spins.

\begin{figure}[t]
\begin{center}
\hspace{-0.1cm}
\includegraphics[width=1\linewidth]{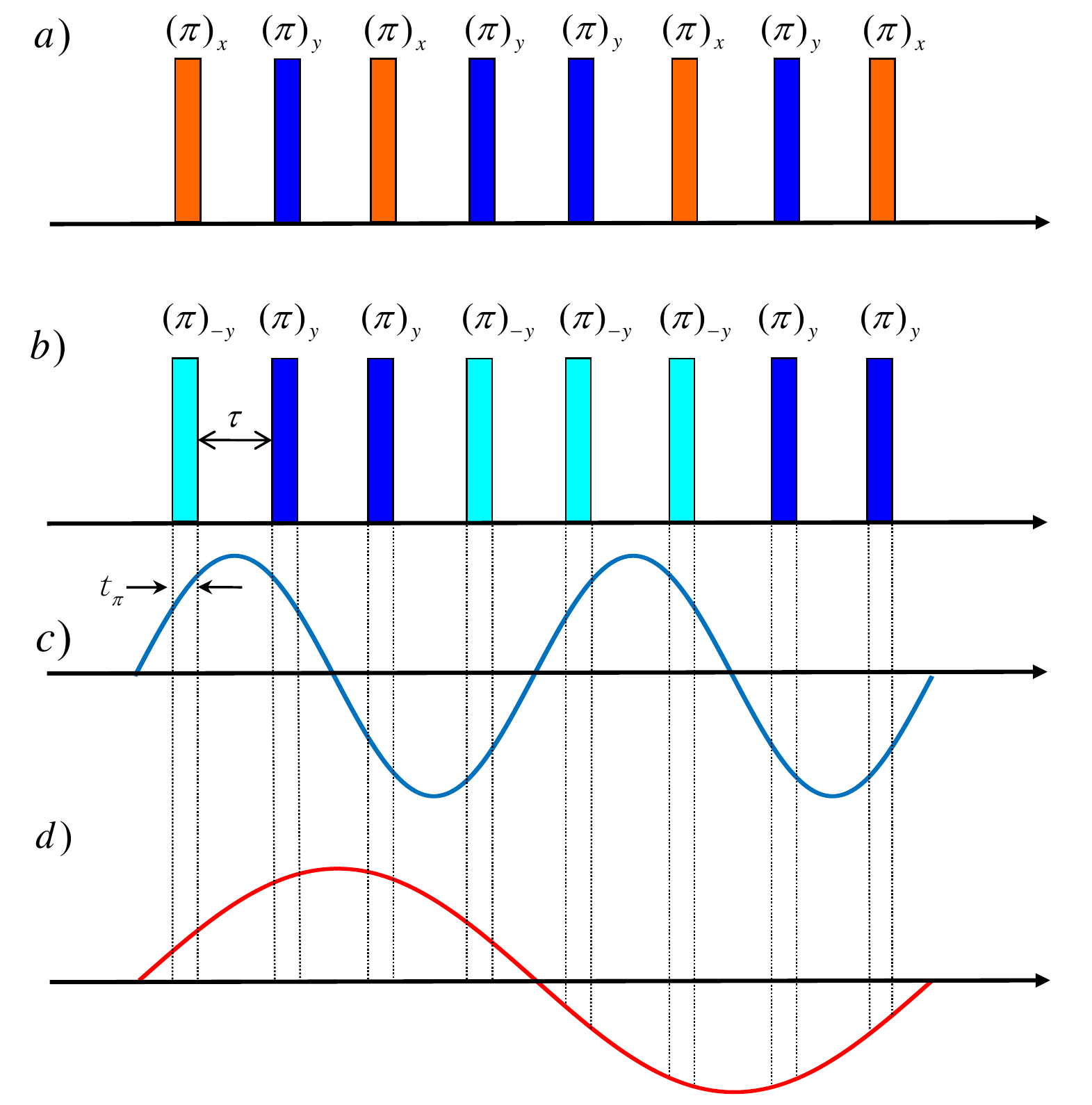}
\end{center}
\caption{(Color online) Robust quantum sensing sequences for single nuclear spin detection.
{\bf (a)} XY8 pulse sequence consisting of repetitive $\blb{\bla{\pi}_{x} \bla{\pi}_{y}
\bla{\pi}_{x} \bla{\pi}_{y}\bla{\pi}_{y} \bla{\pi}_{x}\bla{\pi}_{y} \bla{\pi}_{x}}$ pulses and
{\bf (b)} YY8 pulse sequence consisting of $\blb{\bla{\pi}_{-y} \bla{\pi}_{y} \bla{\pi}_{y} \bla{\pi}_{-y}\bla{\pi}_{-y} \bla{\pi}_{-y}\bla{\pi}_{y} \bla{\pi}_{y}}$ pulses where each bar
refers to a $\pi$ rotation around $\pm \hat{x}, \pm \hat{y}$ axis (with $t_{\pi}$ the pulse duration). {\bf (c-d)} A field
with a frequency of $\omega_{2^{nd}}=\omega_L/2$ and $\omega_{4^{th}}=\omega_c/4$, where $\omega_c=1/(2\tau)$ with $\tau$ the inter pulse time interval, may lead to spurious high-order resonance signal.}
\label{fig:pulses}
\end{figure}

{\it Spurious signal and pulse sequence engineering.---}  Without loss of generality, we
take an NV center spin sensor in diamond to illustrate our idea. We remark that
the present result is also applicable for quantum sensing with the other two-level
physical systems. By choosing e.g. $\{\ket{0},\ket{-1}\}$, one can construct a two-level quantum probe for sensing. By applying a sequence of $N$ equidistant $(\pi)_{\theta_{k}}$ pulses (which represents a $\pi$ rotation around the axis $\cos\theta_{k} \hat{x}+\sin\theta_{k} \hat{y}$) on the NV
center spin with the inter pulse time interval $\tau$ (see Fig.\ref{fig:pulses}),
and assuming that the $(\pi)_{\theta_{k}}$ pulses are instantaneous, the transition
probability from the initial state $\ket{\phi^+}=\bla{\ket{0}+e^{i\phi}\ket{-1}}/\sqrt{2}$ to the
state $\ket{\phi^-}=\bla{\ket{0}-e^{i\phi}\ket{-1}}/\sqrt{2}$ is given by \cite{Lukasz08,Stau13} $P(\tau)
=|\bra{\phi^-} U(N\tau) \ket{\phi^+}|^2$, where $\phi$ is the relative phase in the initial state (i.e. the initial phase of the $\frac{\pi}{2}$ pulse), $U(N\tau)$ is the system evolution operator under the joint action of system Hamiltonian and pulsed action. In the context of nuclear spin detection, nuclear
spins undergo Larmor precession in an external magnetic field, and produce a
contribution to the signal with a characteristic frequency given by the Larmor
frequency. When the pulse interval matches the Larmor frequency of nuclear spin, i.e.
$\omega_{L}=1/(2\tau)$, a resonance signal corresponding to the specific nuclear species
is observed.

The NV center spin state is coherently manipulated by applying microwave driving fields as
\begin{equation}
H_{M}(t)=\frac{\Omega(t)}{2}\cos{(\omega t)}  \left[ \cos\phi(t) {\boldsymbol\sigma}_x+\sin\phi(t) {\boldsymbol\sigma}_y \right],\label{eq:pulse-Ham}
\end{equation}
where ${\boldsymbol\sigma}=({\boldsymbol\sigma}_x,{\boldsymbol\sigma}_y,{\boldsymbol\sigma}_z)$ is the Pauli operator in the subspace spanned by $\{\ket{0},\ket{-1}\}$ that are the eigenstates of $\boldsymbol\sigma_z=\ketbra{-1}{-1}-\ketbra{0}{0}$. A $(\theta)_{\phi}$ rotation around the axis $(\cos\phi,\sin\phi,0)$ realized
in the time duration $\left[t_0,t_0+t_d \right]$ is achieved by choosing
$\Omega(t)=\Omega$ and $\phi(t)=\phi$ for $t\in \left[t_0,t_0+t_d \right]$,
with $t_d=\theta/\Omega$. In the case that the microwave Rabi frequency $\Omega$ is
much larger than the hyperfine interaction, the corresponding evolution operator
of a $(\pi)_{\phi}$ rotation is $U_{\phi}^{\pi}=\cos\phi  {\boldsymbol\sigma}_x+\sin\phi  {\boldsymbol\sigma}_y$.
Under realistic conditions, due to the finite power of the microwave acting on the NV center, the $\pi$-pulses always have a finite width (as opposed to instantaneous pulses), which can lead to spurious harmonic response and thus the possible ambiguity in the identification of nuclei \cite{Loretz15}. Such an effect becomes particularly prominent in single-molecule magnetic resonance spectroscopy as a large number of pulses is required to get a resonance signal from weakly coupled nuclei.

To overcome such a serious obstacle, we need to carefully engineer quantum sensing pulse
sequence. In order to illustrate the basic idea, we first consider a classical
oscillating field $H_{ac}=A_{ac}\sin(\omega_{2^{nd}} t) \frac{{\boldsymbol\sigma}_z}{2}$ with a frequency $\omega_{2^{nd}}
=\omega_{L}/2$, see Fig.\ref{fig:pulses}(c). We prepare the NV spin sensor into
the initial state as $\ket{\phi^+}=\bla{\ket{0}+e^{i\phi}\ket{-1}}/\sqrt{2}$. As the pulse has
a finite width, the corresponding evolution of the spin sensor state is not a
perfect $\pi$-rotation around $\hat{x}$ or $\hat{y}$. Instead, the real
evolution can be described by a set of rotations in the form of $R_{\hat{n}_k}
=\exp{(-i\theta_k \hat{n}_k\cdot  {\boldsymbol{\sigma}}/2)}$, where $\boldsymbol{\sigma}$ is the spin
operator and $\theta_k=\pi+\frac{\pi}{2}\beta_k^2+O(\beta_k^3)$, $\hat{n}_k=(\cos\beta_k\cos\phi_k, \cos\beta_k\sin\phi_k, \sin\beta_k)$ and $\beta_k\approx A_{ac}\sin(\omega t_k)/\Omega\ll1$, $t_k$ is the moment when the pulse is applied. Such imperfection in pulses give rise to spurious resonance signal. For the XY8 pulse sequence, in the scenario that we are interested in, namely $A_{ac}/\Omega\ll 1$, the total evolution for one
cycle is $U_{2^{nd}}=\prod_{k=1}^{8} R_{\hat{n}_k}=(1-8\beta_0^2)\hat{I}+i2\sqrt{2}\beta_0(\sigma_x-\sigma_y)$, where $\beta_0=A_{ac}/\Omega$. This leads
to a $2^{nd}$ harmonic response signal $P_{2^{nd}}=8\left[1-\sin(2\phi)\right]\beta_0^2+O(\beta_0^3)$
\cite{Loretz15,Haase16}, which spuriously indicates a frequency component that is twice of the real oscillating field frequency. Similar analysis shows that the  $4^{th}$ harmonic response signal is $P_{4^{th}}=2(2-\sqrt{2})\left[1+\sin(2\phi))\right]\beta_0^2+O(\beta_0^3)$ \cite{Loretz15,Haase16}. It can be seen that by changing the initial phase $\phi$, $2^{nd}$ and $4^{th}$ harmonics can be suppressed respectively, as $P_{2^{nd}}$ vanishes when $\phi=\pi/4$ while $P_{4^{th}}$ goes to zero when $\phi=3\pi/4$. This may require extra experiment complexity.

We find that a type of MLEV-8 pulse sequence $\blb{\bla{\pi}_{-y}
\bla{\pi}_{y} \bla{\pi}_{y} \bla{\pi}_{-y}\bla{\pi}_{-y} \bla{\pi}_{-y}\bla{\pi}_{y} \bla{\pi}_{y}}^{N}$(for simplicity we call it as YY8 pulse sequence) can suppress both $2^{nd}$ and $4^{th}$ harmonic signals simultaneously and thus reduce the experiment complexity. Using YY8 pulse sequence, the cycle evolution of the NV center spin state for the $2^{nd}$ and $4^{th}$ harmonic response are given by:
\begin{align}
\label{Probability}
P_{2^{nd}}^\prime&=4\blb{1-\cos(2\phi)}\beta_0^2+O(\beta_0^3)\nonumber\\
P_{4^{th}}^\prime&=(2-\sqrt{2})\blb{1-\cos(2\phi)}\beta_0^2+O(\beta_0^3)
\end{align}
It can be seen that the $2^{nd}$ and $4^{th}$ harmonic signal can thus be simultaneously suppressed up to second-order when the initial phase is chosen as $\phi=0$\cite{SI}. We remark that YY8 pulse sequence can suppress the other spurious resonances, including the $(4/5)^{th}$, $(2/3)^{th}$ and $(4/3)^{th}$ response, in contrast with XY8 pulse sequence \cite{SI}.

\begin{figure}[t]
\begin{minipage}{9cm}
\hspace{-0.5cm}
\includegraphics[width=4.6cm]{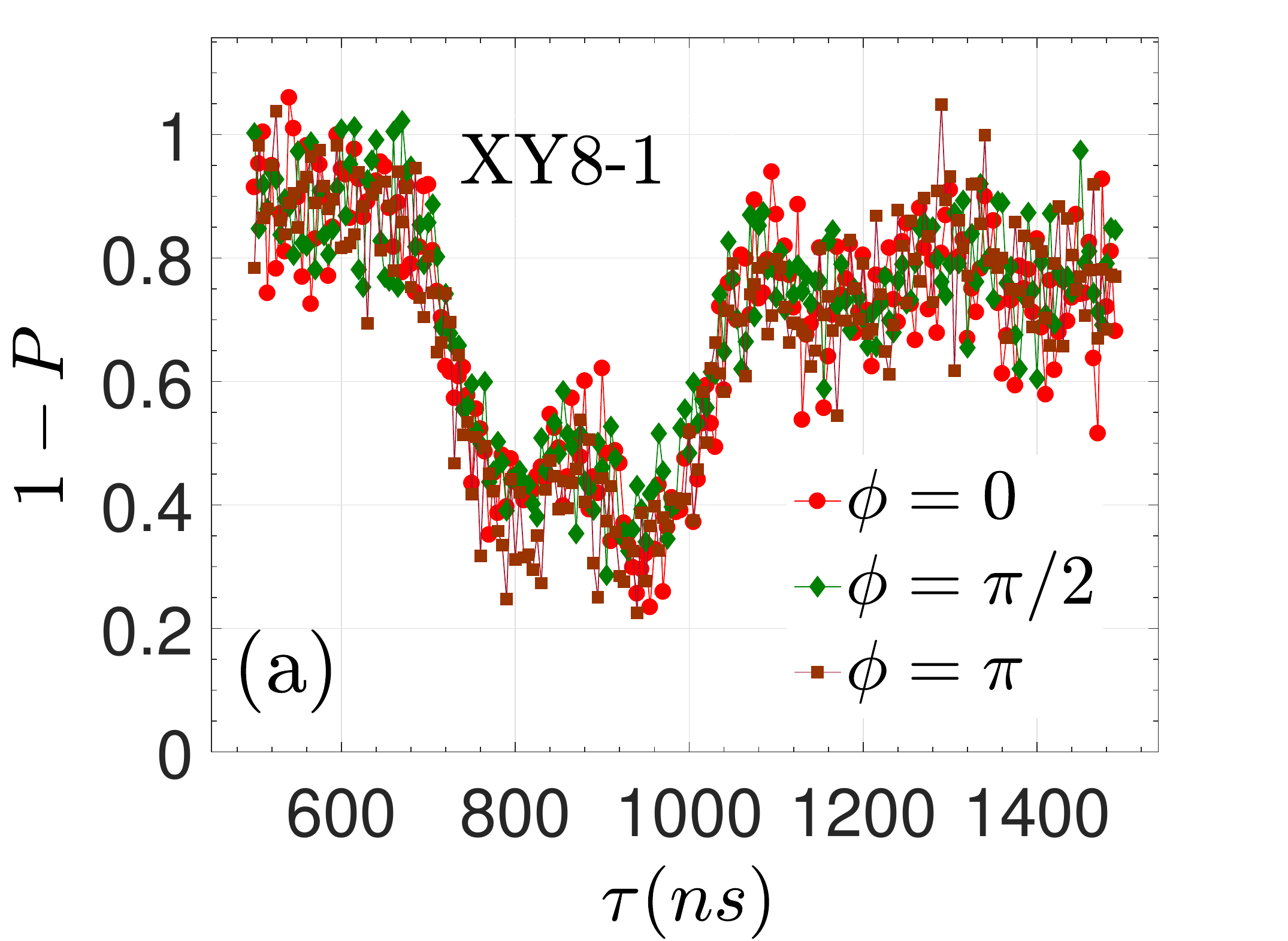}
\hspace{-0.2cm}
\includegraphics[width=4.6cm]{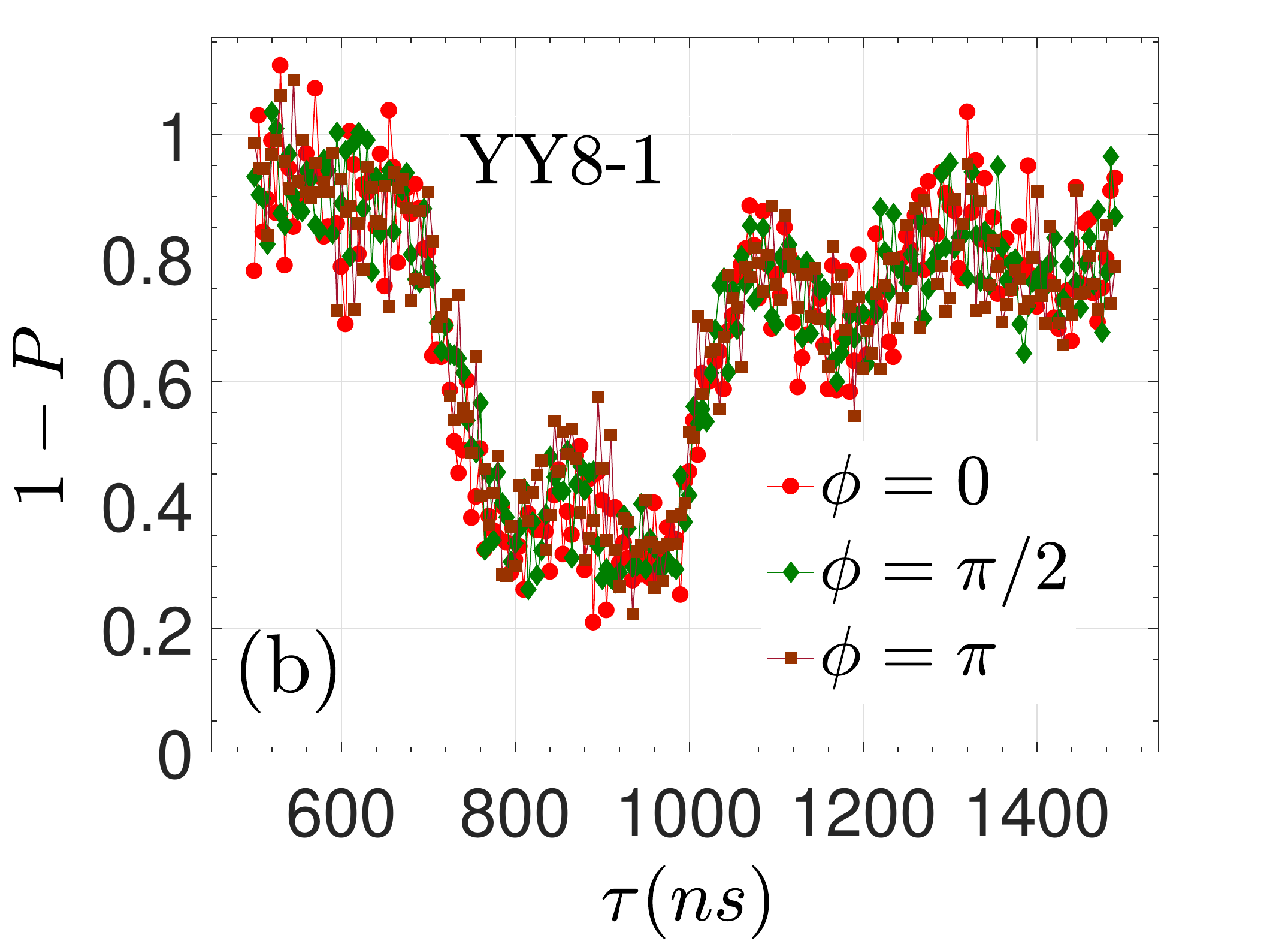}
\end{minipage}
\caption{(Color online) The spectrum signal of the $^{13}$C nuclear spins observed by an NV center using XY8-1 and YY8-1 pulse sequence with different initial phases $\phi=0$ ($\circ$), $\pi/2$ ($\diamond$), $\pi$ ($\square$). The Rabi frequency is $\Omega=(2\pi)25$MHz, the magnetic field is $510$G. }
\label{fig:realsignal}
\end{figure}

{\it Unambiguous nuclear spin detection.---} Using an NV center spin sensor in diamond, we demonstrate experimentally the suppression of spurious high-order response signals in the detection of nuclear spins using engineered quantum sensing pulse sequences. The NV center is a point defect in diamond, which has a triplet
ground state manifold with the spin quantum number $m_s=0,\pm 1$. In NV-based
nuclear magnetic resonance experiments, one applies a magnetic field
$B_z$ along the NV axis that connects the nitrogen atom and the vacancy site,
the NV center spin Hamiltonian is given as follows \cite{Doherty13}
\begin{equation}
H_{0}= D \mathbf{S}_{z}^{2}-\gamma_{e}B_{z}\mathbf{S}_{z}, \label{eq:sys-Ham}
\end{equation}
where the zero-field splitting is $D/2\pi=2.87 \mbox{GHz}$, $\gamma_e$ is the
electron gyromagnetic ratio, $\mathbf{S}$ is the spin-1 operator. As an example, we consider the detection of
$^{13}$C nucleus in bulk diamond. The hyperfine interaction between the NV
center spin and the nuclear spin is described by
\begin{eqnarray}
    H_{I} = S_{z}\sum_{j}\vec{A}_{j}\cdot\vec{I}_j = S_{z}\sum_{j}(a^{j}_{\|}\hat{I_z}^{j} +
    a^{j}_{\bot}\hat{I_x}^{j}),
\end{eqnarray}
where $a_{\|}^{j}$ and $a_{\bot}^{j}$ are the longitudinal and transversal components of the
hyperfine interaction. The hyperfine interaction will cause an effective fluctuating
magnetic field with a characteristic frequency that is equal to the Larmor frequency
of the nuclear spin. The transition probability of the NV spin state at time $T$ is
thus given by $P(T)=\bra{\phi^{-}} \mbox{Tr}_{I} \blb{U(\ketbra{\phi^{+}} {\phi^{+}}\otimes \rho_{N})U^{\dagger}} \ket{\phi^{-}}$, where $U$ is the evolution operator, $\rho_{N}$ represents the initial state of the nuclear spins. The probability that the NV spin remains in the initial state  $P_0=1-P(N\tau)$ is the signal observed in our NV-based nuclear magnetic resonance experiments, which provides us information on the nuclear spin, such as the type of nuclei and the hyperfine interaction with the NV spin sensor.

\begin{figure}[b]
\hspace{-0.2cm}
\includegraphics[width=9cm]{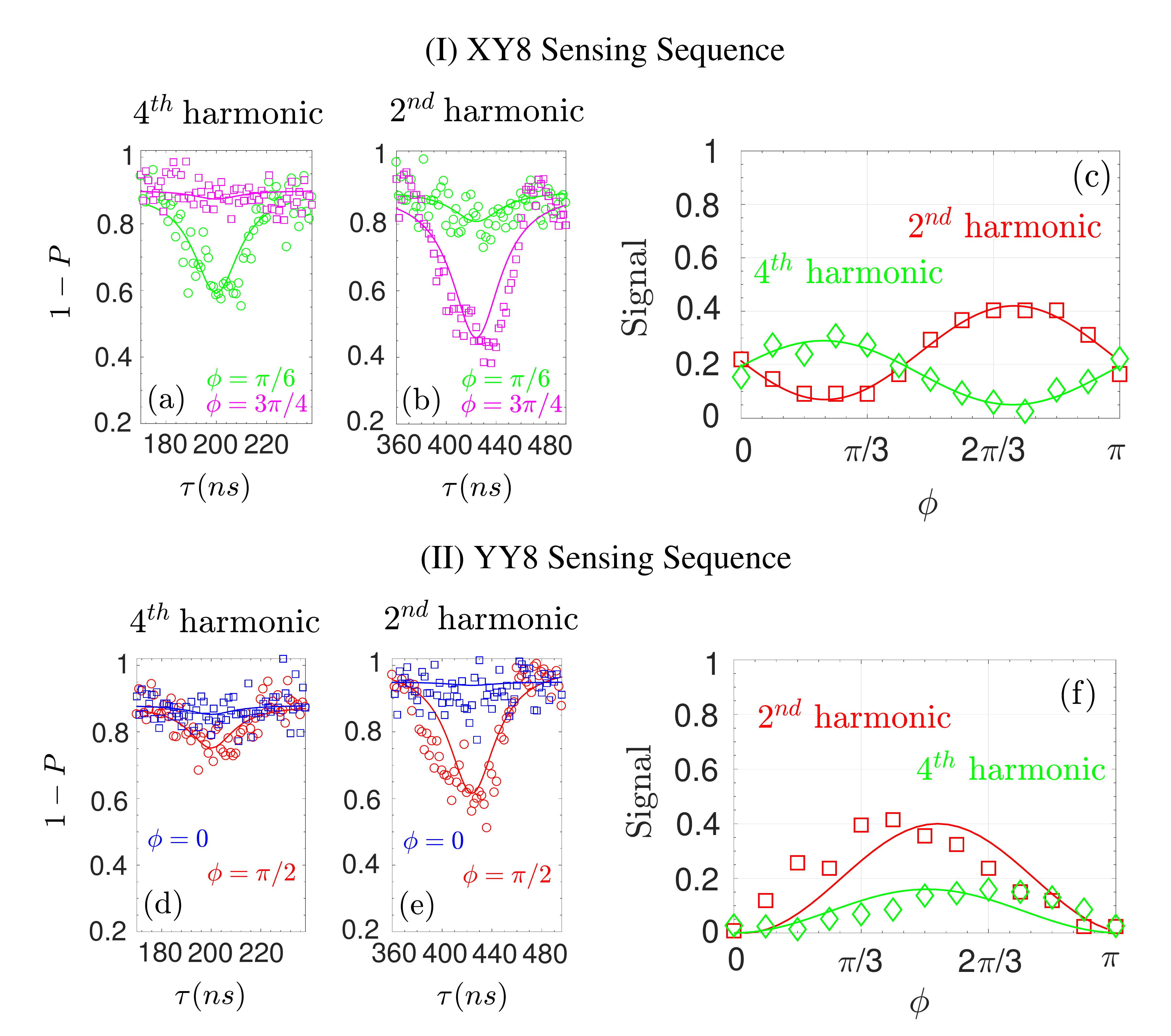}
  \caption{(Color online) Suppression of spurious high-order resonance signals using engineered quantum sensing pulse sequences. ({\bf a-b}) The 2$^{nd}$ and 4$^{th}$ harmonic response using XY8-6 and XY8-12 pulse sequence with the initial phase $\phi=\pi/6$ ($\circ$) and $3\pi/4$ ($\square$) respectively. ({\bf c}) The amplitude of the 2$^{nd}$ ($\square$) and 4$^{th}$ ($\diamond$) harmonic signal (i.e. the resonance contrast) as a function of the initial phase using XY8-6 and XY8-12 pulse sequence. ({\bf d-e}) The 2$^{nd}$ and 4$^{th}$ harmonic response using YY8-6 and YY8-12 pulse sequence with the initial phase $\phi=0$ ($\square$) and $\pi/2$ ($\circ$) respectively. ({\bf f}) The amplitude of the 2$^{nd}$ ($\square$) and 4$^{th}$ ($\diamond$) harmonic signal as a function of the initial phase using YY8-6 and YY8-12 pulse sequence. The Rabi frequency is $\Omega=(2\pi)25$ MHz. Both spurious harmonics can be suppressed simultaneously for YY8 pulse sequences which contrasts with XY8 pulse sequences.}\label{fig:2nd}
\end{figure}

In  our experiment, we polarize and read out the NV center spin in type IIa diamond (with natural abundance of $^{13}$C) using optical techniques. The microwave control pulses are generated using an arbitrary wave generator (Tektronix AWG70002A) and is amplified by an 80W microwave amplifier using a same channel. This allows us to avoid the amplitude imbalance between pulses, which otherwise would affect the robustness of YY8 pulse sequence \cite{Levitt83}. Therefore, YY8 pulses may be more robust than XY8 sequence as is also evidenced by our numerical simulation \cite{SI}. The amplitude and phase are tuned to implement the engineered quantum sensing pulse sequences. We apply a magnetic field (510G) along the NV axis, under which condition the associated $^{14}$N can be polarized to improve the sensing signal, and use $\ket{m_s=0}$ and $\ket{m_s=-1}$ as quantum probe in our sensing experiments. We first measure the magnetic resonance signal of $^{13}$C nuclear spin using XY8 and YY8 pulse sequences, see Fig.\ref{fig:realsignal}. It can be seen that the signal is almost independent on the initial phase, and both XY8 and YY8 pulse sequences demonstrate similar performance.

\begin{figure}[t]
\begin{minipage}{9cm}
\hspace{-0.5cm}
\includegraphics[width=9cm]{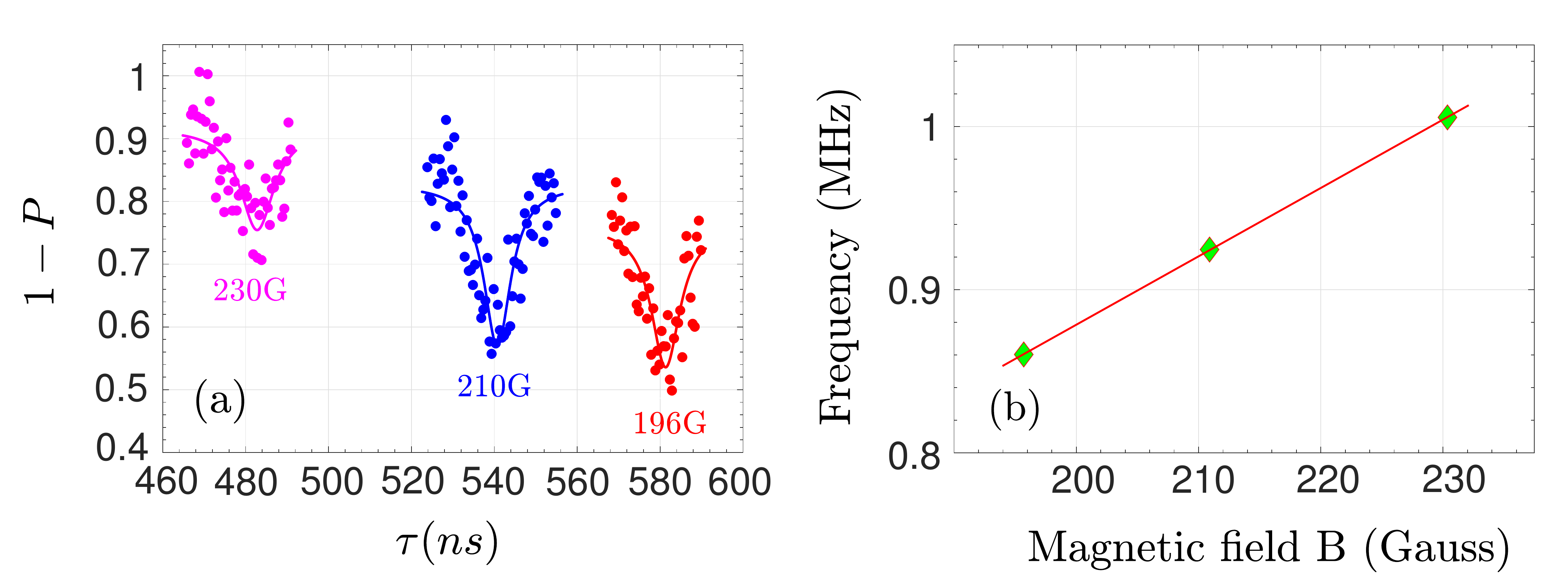}
\end{minipage}
  \caption{(Color online) Detection of $^1$H magnetic resonance signal using YY8 pulse sequence. ({\bf a}) The spectrum as measured by a shallow NV center spin for various magnetic field. ({\bf b}) The frequency of peak response as a function of the magnetic field $B$. The red line has a slope $4.20\pm 0.13$ kHz/G that gives an estimated gyromagnetic ratio of $^1$H. The Rabi frequency is $\Omega=(2\pi)25$ MHz.}\label{fig:H_signal}
\end{figure}

We further test the suppression of spurious high-order resonance signals using engineered pulse sequences. In Fig.\ref{fig:2nd}, we plot the dependence of the $2^{nd}$ and $4^{th}$ harmonic signal on the initial phase $\phi$ using the YY8 pulse (d-f) sequence as compared with the XY8 sequence (a-c). It can be seen that the spurious high-order resonance signals can be suppressed by choosing approximate initial phases of XY8 and YY8 pulse sequences, in which YY8 pulse sequence demonstrates better performance over XY8 pulse sequence. By choosing $\phi=0$, both $2^{nd}$ and $4^{th}$ harmonics are suppressed simultaneously using YY8 pulse sequence, see Fig.\ref{fig:2nd} (d-f), while this cannot be achieved for any choice of $\phi$ for the XY8 sequence, see Fig.\ref{fig:2nd}(a-c). The amplitude of  spurious high-order resonance signals are also relatively smaller using YY8 pulse sequence even when the initial phase is not perfect. These advantageous features of YY8 pulse sequence allows for the efficient unambiguous identification of nuclear spins, in particular it implies that YY8 pulse sequence can unambiguously identify the signal from $^1$H, while avoiding the possible confusion with the $4^{th}$ harmonic signal of $^{13}$C with no extra experiment requirement as compared with the already widely used XY8 pulse sequence.

We then use YY8 pulse sequence with the initial phase $\phi=0$ to detect the magnetic resonance signal of $^1$H in oil on diamond surface. The shallowed implanted NV centers are created by a $2.5$ keV N$^+$ ions into type IIa diamond. The Ramsey and spin echo measurement indicates that the NV center shows $T_2^*\approx 300$ns and $T_2\approx5\mu $s. In Fig.\ref{fig:H_signal}(a), we measure the spectrum recorded by YY8 pulse sequence at different magnetic fields. The scaling of the peak frequency with the magnetic field is shown in Fig.\ref{fig:H_signal}(b). The linear slope of the frequency $4.20\pm 0.13$kHz/G fits well with the gyromagnetic ratio of $^1$H ($4.258$ kHz/G). As we show above, the YY8 pulse sequence can effectively suppress the high-order spurious response, and therefore our measurement provides a way to detect $^1$H unambiguously. As $^1$H magnetic resonance spectroscopy is very important in various scenarios, the present engineered quantum sensing pulse sequence, particularly YY8 pulse sequence, may find useful applications in single-molecule magnetic resonance spectroscopy using quantum spin sensor.

{\it Conclusions---} We implement engineered quantum sensing pulse sequences that
suppress spurious higher order resonances, a feature that will find application in
unambiguous nuclear spin detection and more generally the identification of frequency components of
externally applied signals. Specifically, this eliminates the ambiguity in the nuclear
magnetic resonance signal of single nuclear spin detection that is inherent to
pulsed schemes due to the duration of individual pulses. The YY8 pulse sequences
shows better performance in suppressing spurious response and are even more robust
against the pulse errors than the widely used XY8 pulse sequences (e.g amplitude error).
The simplicity and advantageous features of this type of pulse sequences make them
an appealing choice for the applications in nuclear spin detection as well as the
quantum control of nuclear spin registers.

{\em Acknowledgements --} We thank Jan Haase and Benedikt Tratzmiller for helpful comments and suggestions, and Michael Ferner and Manfred B{\"u}rzele for technical assistance. J.-M.C is supported by National Natural Science Foundation of China (11574103, 11690030, 11690032).  M. B. P. is supported by the DFG (FOR1493 and SFB TR21), the EU STREPs EQUAM, DIADEMS and HYPERDIAMOND, and the ERC Synergy grant BioQ. C. M., J. L., N. T., B. N., L. P. M., F. J. acknowledge support from DFG (FOR 1493, SFB TR21, SPP 1923), VW Stiftung, BMBF, ERC, EU (DIADEMS), BW Stiftung, Ministry of Science and Arts, Center for Integrated quantum science and technology (IQST).


\begin{thebibliography}{99}

\bibitem{Rugar04} D. Rugar, R. Budakian, H. J. Mamin, and B. W. Chui, {\it Nature} {\bf 430}, 329 (2004)

\bibitem{Hue13} S. F. Huelga and M. B. Plenio. {\it Contemp. Phys.} {\bf 54}, 181-207 (2013).

\bibitem{Lambert13} N. Lambert, Y.-N. Chen, Y.-C. Cheng, C.-M. Li, G.-Y. Chen and F. Nori, {\it Nature Physics} {\bf 9}, 10 (2013).


\bibitem{Vincent12} R. Vincent, S. Klyatskaya, M. Ruben, W. Wernsdorfer, and F. Balestro, {\it Nature} {\bf 488}, 357 (2012).

\bibitem{Pla13} J. J. Pla, K. Y. Tan, J. P. Dehollain, W. H. Lim, J. J. L. Morton, F. A. Zwanenburg, D. N. Jamieson, A. S. Dzurak, and A. Morello, {\it Nature} {\bf 496}, 334 (2013).

\bibitem{Tam12} T. H. Taminiau, J. J. T. Wagenaar, T. van der Sar, F. Jelezko, V. V. Dobrovitski, and R. Hanson, {\it Phys. Rev. Lett.} {\bf 109}, 137602 (2012).

\bibitem{Kol12}  S. Kolkowitz, Q. P. Unterreithmeier, S. D. Bennett, and M. D. Lukin, {\it Phys. Rev. Lett.} {\bf  109}, 137601 (2012).

\bibitem{Zhao12}  N. Zhao, J. Honert, B. Schmid, M. Klas, J. Isoya, M. Markham, D. Twitchen, F. Jelezko, R.-B. Liu, H. Fedder, and J. Wrachtrup, {\it Nature Nanotechnology} {\bf 7}, 657 (2012).

\bibitem{Mamin13} H. J. Mamin, M. Kim, M. H. Sherwood, C. T. Rettner, K. Ohno, D. D. Awschalom, and D. Rugar, {\it Science} {\bf 339}, 557 (2013).

\bibitem{Stau13} T. Staudacher, F. Shi, S. Pezzagna, J. Meijer, J. Du, C. A. Meriles, F. Reinhard, and J. Wrachtrup, {\it
Science} {\bf 339}, 561 (2013).

\bibitem{Muller14} C. M{\"u}ller, X. Kong, J.-M. Cai, K. Melentijevic, A. Stacey,
M. Markham, J. Isoya, S. Pezzagna, J. Meijer, J.-F. Du, M. B. Plenio, B. Naydenov,
L. P. McGuinness and F. Jelezko. {\it Nature Communications} {\bf 5}, 4703 (2014).

\bibitem{Sus14} A. O. Sushkov, I. Lovchinsky, N. Chisholm, R. L. Walsworth, H. Park, and M. D. Lukin, {\it Phys. Rev. Lett.}
{\bf 113}, 197601 (2014).

\bibitem{Shi15} F.-Z. Shi, Q. Zhang, P.-F. Wang. H.-B. Sun, J.-R. Wang, X. Rong, M. Chen, C.-Y Ju, F. Reinhard, H.-W. Chen, J. Wrachtrup, J.-F. Wang, J.-F. Du, {\it Science} {\bf 347}, 1135 (2015).

\bibitem{Per14} V. S. Perunicic, L. T. Hall, D. A. Simpson, C. D. Hill, L. C. L. Hollenberg, {\it Phys. Rev. B} {\bf 89}, 054432 (2014).

\bibitem{Doherty13} M. W. Doherty, N. B. Manson, P. Delaney, F. Jelezko,
J. Wrachtrup and L. C.L. Hollenberg, Phys. Reports \textbf{528}, 1
(2013).

\bibitem{Wu16} Y. Wu, F. Jelezko, M. B. Plenio, and T. Weil.  {\it Angewandte Chemie- International Edition Minireview} {\bf 55}, 6586 (2016).

\bibitem{Schir14} R. Schirhagl, K. Chang, M. Loretz, and C. L. Degen, {\it Annu. Rev. Phys. Chem.} {\bf 65} 83 (2014).

\bibitem{Erm13} A. Ermakova, G. Pramanik, J.-M. Cai, G. Algara-Siller, U. Kaiser, T. Weil, H. C. Chang, L.P . McGuinness, M. B. Plenio, B. Naydenov and F. Jelezko, {\it Nano Letters} {\bf 13}, 3305 (2013).

\bibitem{Jiang09} L. Jiang, J. S. Hodges, J. R. Maze, P. Maurer, J. M. Taylor, D. G. Cory,
P. R. Hemmer, R. L. Walsworth, A. Yacoby, A. S. Zibrov, and M. D. Lukin, {\it Science} {\bf 326}, 267 (2009).

\bibitem{Maurer12} P. C. Maurer, G. Kucsko, C. Latta, L. Jiang, N. Y. Yao, S. D. Bennett, F. Pastawski, D. Hunger, N. Chisholm, M. Markham, D. J. Twitchen, J. I. Cirac, and M. D. Lukin, {\it Science} {\bf 336}, 1283 (2012).

\bibitem{Waldherr14} G. Waldherr, Y. Wang, S. Zaiser, M. Jamali, T. SchulteHerbrueggen,
H. Abe, T. Ohshima, J. Isoya, P. Neumann, and J. Wrachtrup, {\it Nature} {\bf 567}, 204 (2014).

\bibitem{Cai13} J.-M. Cai, A. Retzker, F. Jelezko, and M. B. Plenio, {\it Nature Physics} {\bf 9}, 168 (2013).

\bibitem{CaiNJP13} J.-M. Cai, F. Jelezko, M. B. Plenio, A. Retzker, {\it New J. Phys.} {\bf 15}, 013020 (2013).

\bibitem{Lukasz08} L. Cywi\'nski, R.M. Lutchyn, C.P. Nave, and S. Das Sarma,
{\it Phys. Rev. B} {\bf 77}, 174509 (2008).

\bibitem{Zhao14} N. Zhao, J. Wrachtrup, R.-B. Liu, {\it Phys. Rev. A} {\bf 90}, 032319 (2014).

\bibitem{Ma15} W.-C. Ma, F.-Z. Shi, K.-B. Xu, P.-F. Wang, X.-K. Xu, X. Rong, C.-Y. Ju, C.-K. Duan, N. Zhao, and J.-F. Du, {\it Phys. Rev. A} {\bf 92}, 033418 (2015).

\bibitem{Greiner15} J. N. Greiner, D. D. Bhaktavatsala Rao, P. Neumann, J. Wrachtrup, arXiv:1507.05457.

\bibitem{Albrecht15} A. Albrecht and M. B. Plenio, {\it Phys. Rev. A} {\bf 92}, 022340 (2015)

\bibitem{Cas15} J. Casanova, Z.-Y. Wang, J. F. Haase, and M. B. Plenio, {\it Phys. Rev. A} {\bf 92}, 042304 (2015).

\bibitem{Wang15} Z.-Y. Wang, J. F. Haase, J. Casanova, and M. B. Plenio, {\it Phys. Rev. B} {\bf 93}, 174104 (2016).

\bibitem{Gullion90} T. Gullion, D. B. Baker, and M. S. Conradi, {\it J. Magn. Reson.} {\bf 89}, 479 (1990).

\bibitem{Ryan10} C. A. Ryan, J. S. Hodges, and D. G. Cory, {\it Phys. Rev. Lett.} {\bf 105}, 200402 (2010).

\bibitem{Cai12} J.-M. Cai, B. Naydenov, R. Pfeiffer, L. McGuinness, K. D. Jahnke, F. Jelezko, M. B. Plenio and A. Retzker, {\it New J. Phys.} {\bf 14}, 113023 (2012).

\bibitem{Loretz15} M. Loretz, J. M. Boss, T. Rosskopf, H. J. Mamin, D. Rugar, and C. L. Degen, {\it Phys. Rev. X} {\bf 5}, 021009 (2015).

\bibitem{Haase16}J. F. Haase, Z.-Y. Wang, J. Casanova, and M. B. Plenio, {\it Phys. Rev. A} {\bf 94}, 032322 (2016)

\bibitem{Sta15} T. Staudacher, N. Raatz, S. Pezzagna, J. Meijer, F. Reinhard, C. A. Meriles, J. Wrachtrup, {\it Nat. Commun.} {\bf 6}, 8527 (2015).

\bibitem{Lev83} M. H. Levitt, R. Freeman, T. Frenkiel, in {\it Advances in Magnetic Resonance} (J. S. Waugh, Ed.), Academic Press, New York, 1983.


\bibitem{SI} See supplementary material for the calculation details, which includes Refs. \cite{Solomon59,Aiello13,Mk14,Mk15,Lange10,Cabrera18,Zhihao}


\bibitem{Levitt83} M. H. Levitt, R. Freeman, and T. Frenkiel,  {\it Advances in Magnetic Resonance} (J. S. Waugh, Ed.), Vol. 11, Academic Press, New York (1983).

\bibitem{Lange10} G. de Lange, Z. H. Wang, D. Rist$\grave{\mbox{e}}$, V. V. Dobrovitski,
R. Hanson, {\it Science} {\bf 10}, 1126 (2010).

\bibitem{Zhihao} Z.-H. Xiao , L.-W. He , and W.-G. Wang.{\it Phys.Rev.A}{\bf 83}, 032322 (2011).

\bibitem{Solomon59} I. Solomon, {\it Phys. Rev. Lett.} {\bf 2}, 301 (1959).

\bibitem{Aiello13} C. D. Aiello, M. Hirose, P. Cappellaro, {\it Nature Communications} {\bf 4}, 1419 (2013).

\bibitem{Mk14} V. V. Mkhitaryan and V. V. Dobrovitski, {\it Phys. Rev. B} {\bf  89}, 224402 (2014).

\bibitem{Mk15} V. V. Mkhitaryan, F. Jelezko, V. V. Dobrovitski, {\it Scientific Reports} {\bf 5}, 15402 (2015).

\bibitem{Cabrera18} R. Cabrera, W.E. Baylis, {\it Phys. Lett. A.} {\bf 368}, 25 (2007).




\end{thebibliography}
\end{document}